# Interface control of ferroelectricity in a SrRuO$_3$/BaTiO$_3$/SrRuO$_3$ capacitor and its critical thickness

*Yeong Jae Shin, Yoonkoo Kim, Sung-Jin Kang, Ho-Hyun Nahm, Pattukkannu Murugavel, Jeong Rae Kim, Myung Rae Cho, Lingfei Wang, Sang Mo Yang, Jong-Gul Yoon, Jin-Seok Chung, Miyoung Kim, Hua Zhou, Seo Hyoung Chang\*, Tae Won Noh\**

Y. J. Shin, Dr. S.-J. Kang, Dr. H.-H. Nahm, J. R. Kim, Dr. M. R. Cho, Dr. L. Wang, Prof. T. W. Noh
Center for Correlated Electron Systems, Institute for Basic Science (IBS), Seoul 08826, Republic of Korea
Department of Physics and Astronomy, Seoul National University, Seoul 08826, Republic of Korea
E-mail: twnoh@snu.ac.kr
Y. Kim, Prof. M. Kim
Department of Materials Science and Engineering and Research Institute of Advanced Materials, Seoul National University, Seoul 08826, Republic of Korea
Prof. P. Murugavel
Department of Physics, Indian Institute of Technology Madras, Chennai 600036, India
Prof. S. M. Yang
Department of Physics, Sookmyung Women's University, Seoul 04310, Republic of Korea
Prof. J.-G. Yoon
Department of Physics, University of Suwon, Hwaseong, Gyunggi-do 18323, Republic of Korea
Prof. J.-S. Chung
Department of Physics, Soongsil University, Seoul 06978, Republic of Korea
Dr. H. Zhou
Advanced Photon Source, Argonne National Laboratory, Lemont, Illinois 60439, United States
Prof. S. H. Chang
Department of Physics, Pukyong National University, Busan 48513, Republic of Korea
E-mail: cshyoung@pknu.ac.kr

The atomic-scale synthesis of artificial oxide heterostructures offers new opportunities to create novel states that do not occur in nature. The main challenge related to synthesizing these structures is obtaining atomically sharp interfaces with designed termination sequences.[1,2] Here, we demonstrate that the oxygen pressure ($P_{O2}$) during growth plays an important role in controlling the interfacial terminations of SrRuO$_3$/BaTiO$_3$/SrRuO$_3$ (SRO/BTO/SRO) ferroelectric capacitors. The SRO/BTO/SRO heterostructures were grown by the pulsed laser deposition (PLD) method. The top SRO/BTO interface grown at high $P_{O2}$ (around 150 mTorr) usually exhibited a mixture of RuO$_2$-BaO and SrO-TiO$_2$ terminations. By reducing $P_{O2}$, we obtained atomically sharp SRO/BTO top interfaces with uniform SrO-TiO$_2$ termination. Using capacitor devices with symmetric and uniform interfacial termination, we were able to demonstrate for the first time that the ferroelectric (FE) critical thickness can reach the theoretical limit of 3.5 unit cells (u.c.).[3]

Oxide heterostructures have attracted significant research interest due to the discovery of novel emergent phenomena.[2,4–13] As a result, heteroepitaxy growth techniques have advanced significantly during the last two decades. However, the task of realizing high-quality oxide heterostructures with atomically sharp interfaces is still challenging, even for simple perovskite oxides. For instance, it has been widely accepted that each maximum of a reflective high-energy electron diffraction (RHEED) intensity profile during PLD growth corresponds to the growth of one unit cell. The perovskite stacking sequence is therefore expected to be preserved during stoichiometric deposition (i.e. a *unit-cell-by-unit-cell* growth scheme). However, recent studies have shown that RHEED oscillations alone do not provide sufficient information to confirm the intended atomic structure of the surface.[14–16] During PLD growth, numerous growth variables, including the chemical stabilities and mobilities of surface adatoms, can be altered by the growth conditions. This alternation may in turn give rise to uncontrollable changes of surface termination.[15,16] Up to this point, there have been few investigations on how the key growth parameters, e.g. $P_{O2}$ and temperature, affect the surface termination.

Ultrathin FE heterostructures are model systems in which the physical properties are highly dependent on the interfacial structure. As the FE film thickness decreases, the FE polarization decreases monotonically and finally disappears at a critical thickness. According to theoretical predictions, the prototypical all-oxide SRO/BTO/SRO heterostructure should have a critical thickness between 3.5 – 6.5 u.c..[3,17] Experimental critical thicknesses are, however, significantly larger. Recent studies on SRO/BTO/SRO capacitors have shown that the lack of atomic scale interface control is the main obstacle to obtaining FE critical thicknesses approaching the theoretical limit.[18,19] The experimental structures feature mixed terminations and the resulting pinned dipoles notably degrade the FE polarization stability, which results in the suppression of ferroelectric behavior below about 20 u.c..[19]

Here, we demonstrate the critical role of $P_{O2}$ in determining the interface termination. Using scanning transmission electron microscopy (STEM) and x-ray diffraction, the interfaces of SRO/BTO heterostructures were studied for various $P_{O2}$ = 5 and 150 mTorr. In the 150 mTorr case, we found a mixed interfacial termination. However, for $P_{O2}$ = 5 mTorr, a singly terminated interface was observed. Our density functional theory (DFT) calculations also support that the stability of the possible interface atomic structures are significantly affected by changing $P_{O2}$. Capacitor devices grown at $P_{O2}$ = 5 mTorr with well-controlled interfaces show FE critical thicknesses as low as 3.5 u.c., in agreement with theory.[3]

The SRO/BTO/SRO capacitor can have two possible atomic arrangements, as illustrated in **Scheme 1**. Since $RuO_2$ is thermally unstable during high-temperature deposition, the bottom SRO layer is always SrO terminated.[16] The BTO layer, therefore, starts with $TiO_2$. Following the commonly assumed *unit-cell-by-unit-cell growth* mode, the growth of the BTO layer should end with a top BaO termination. As a result, the top interface should have the sequence BaO-$RuO_2$, which produces an asymmetric capacitor configuration (Scheme 1b). On the other hand, the FE critical thickness calculations have been performed only for ideal symmetric configurations with TiO$_2$ termination on both top and bottom BTO surfaces (Scheme 1a).[3,17] Obtaining such a symmetric termination configuration is essential to test the theoretically predicted value of the FE critical thickness.

Fully strained SRO/BTO/SRO heterostructures were fabricated using PLD on atomically smooth TiO$_2$-terminated SrTiO$_3$(001) substrates. The thicknesses of the bottom and top SRO electrodes were fixed at 20 nm. During the BTO growth, $P_{O2}$ was set to either 5 or 150 mTorr. The BTO film thickness ($t_{BTO}$) was controlled by monitoring the high pressure RHEED intensity oscillations. In the ideal layer-by-layer growth mode, the number ($n$) of RHEED oscillations signifies a BTO layer with thickness $t_{BTO} = n$ u.c.. However, by using STEM, we found that the films grown at $P_{O2}$ = 5 mTorr end with a half-unit-cell BTO layer (Figure S1, Supporting Information). In this case, we denote the film thickness by $t_{BTO} = (n-0.5)$ u.c.. Using x-ray diffraction (Figure S2, Supporting Information), we exclude the formation of secondary phases for both growth conditions ($P_{O2}$ = 150 and 5 mTorr).

For $P_{O2}$ = 150 mTorr, we found that two types of termination sequence coexist at the top SRO/BTO interface. **Figure 1**a shows a cross-sectional high-angle annular dark field (HAADF) image from STEM, viewed along the [100] zone axis. The HAADF image displays an atomically sharp BTO/SRO bottom interface with the expected TiO$_2$-SrO termination sequence. By contrast, the SRO/BTO top interface termination is highly inhomogeneous. Figure 1b and 1c presents the magnified HAADF images for two regions marked in Figure 1a (dashed boxes I and II). The intensity profiles along the B-site cations are also plotted in the right panels. Note that the intensity of the Ti peaks is lower than those of the Ru peaks due to the smaller atomic number. Four (three) TiO$_2$ layers imply that the BTO film thickness is 3.5 u.c. (3 u.c.), and the resulting SrO-TiO$_2$ (RuO$_2$-BaO) interface will produce a symmetric (asymmetric) capacitor configuration. The mixed termination sequence of the top surface was confirmed by STEM images of different regions of the sample (Figure S3 in the Supporting Information).

In the SRO/BTO/SRO capacitor with mixed interfacial termination, the local FE response varies significantly. Using the electron-beam lithography technique, we fabricated a 5 × 5 μm² square-shaped capacitor (see Experimental Section). We then chose 10×10 grid positions on the SRO top electrode layer and characterized the local FE responses at each grid point by piezoresponse force microscopy (PFM) (See Figure S4 in the Supporting Information for details).[20] As shown in Figure 1d, we can observe two typical piezoresponse signals. In region (i) (left panel), the PFM phase-voltage curve shows a fully saturated hysteresis and the PFM amplitude-voltage curve has a nearly symmetric butterfly shape, indicating robust and switchable ferroelectricity. In region (ii) (right panel), by contrast, the PFM amplitude and phase show FE polarization switching characteristics only for the downward polarization state (pointing towards the bottom SRO). This result implies that the FE polarization in the region (ii) is strongly pinned along one direction. The existence of pinned dipoles in the mixed terminated SRO/BTO/SRO capacitor was also proposed by earlier works.[18,19,21]

In order to map the spatial variation of the ferroelectricity, we calculated the ratio between the PFM amplitudes at ±6 V ($A_{+\text{max}}/A_{-\text{max}}$) at each sampling point. As shown in the central panel of Figure 1d, for approximately 50% of points the $A_{+\text{max}}/A_{-\text{max}}$ ratio is close to 1.0, signifying switchable FE response [region (i)]. For the remaining points, $A_{+\text{max}}/A_{-\text{max}}$ is much higher than 1.0, signifying the response of pinned dipoles [region (ii)]. It is therefore difficult to test the FE critical thickness experimentally with the SRO/BTO/SRO capacitors grown at $P_{\text{O2}}$ = 150 mTorr.

In contrast, the heterostructure grown at $P_{\text{O2}}$ = 5 mTorr has a uniform BTO top interface and a symmetric termination sequence (Scheme 1a). As shown in **Figure 2**a and 2b, the high-resolution HAADF images reveal atomically sharp interfaces formed over a lateral range of 40 nm. The BTO layer is uniformly TiO$_2$-terminated for both the top and bottom interfaces. The intensity profile along the solid box in Figure 2b also confirms the symmetric SrO-TiO$_2$ interface termination sequences (Figure 2c). In this sample with alternatively stacked 3 BaO and 4 TiO$_2$ layers, we denote the $t_{BTO}$ as 3.5 u.c (Figure 2d).

Considering that the STEM images only provide local structural information, we also performed surface x-ray scattering measurements and a coherent Bragg rod analysis (COBRA) to further confirm the uniformity of the BTO interfaces.[22] The sample used for this measurement contains a 5 mTorr grown BTO layer (3.5 u.c.) sandwiched between two SRO layers (3 u.c. each). The COBRA electron density mapping (Figure 2e) indicates that the symmetric SrO-TiO$_2$ interface termination occurs uniformly over the size of the x-ray beam spot, which is typically 30 μm in diameter. These structural investigations clearly indicate that the heterostructure grown with $P_{O2}$ = 5 mTorr has a uniformly TiO$_2$-terminated interface.

We now turn to possible interpretations of the observed $P_{O2}$-dependent termination change. We first checked the cation stoichiometry of the BTO films. Some earlier studies showed that different background pressures could change the composition or stoichiometry of BTO during the growth.[23–25] However, STEM-energy dispersive spectroscopy confirms that our films grown at different $P_{O2}$ are close to stoichiometric with the Ti/Ba ratio very close to one (Figure S5, Supporting information). To check the oxygen stoichiometry, we evaluated the lattice constant of the BTO layer with $P_{O2}$ = 5 mTorr from the COBRA electron density map. The measured c-axis lattice constant was quite consistent with that of a fully strained BTO film on a SrTiO$_3$ (001) substrate. This indicates that there is no significant amount of oxygen vacancies in our BTO layer even with a lower $P_{O2}$ condition. Therefore, we exclude the possibility that deviations in film stoichiometry play a significant role in affecting the composition of the BTO surface and interface termination.

Another possible interpretation is that the $P_{O2}$ dependent termination could be related to the stability of BaO and TiO$_2$ terminations during the growth. The stability of each component can play an important factor in determining the interface terminations. A typical example is epitaxial SRO film growth on SrTiO$_3$ (001) substrates. It is already widely known that termination conversion can occur during the growth of epitaxial SRO films.[16,26–28] The highly volatile nature of RuO$_2$ makes the RuO$_2$-termination unstable during the high-temperature growth, leading to a conversion to SrO-termination. In the BTO case, the stabilities of possible surface structures were reported to be strongly dependent on the background oxygen pressure.[29,30] A change in relative stability may also act as a driving force for forming different surface terminations of BTO. To shed light on this idea, we performed DFT calculations on the Gibbs free energies of BaO- and TiO$_2$-terminated BTO surface (Section VI, Supporting Information). The DFT calculations showed that the TiO$_2$ termination becomes more stable with decreasing $P_{O2}$ (Figure S6, Supporting Information). This is qualitatively consistent with our experimental observations. The change of stability may imply a different bonding strength for each termination which may in turn affect the actual kinetic growth process of PLD and ultimately the termination. Nonetheless, the non-equilibrium nature of PLD makes connecting the DFT results with our experiment challenging. Additional factors that could also affect the surface termination include the density of excited arrived species and surface/arrived species interactions. Further study of the exact mechanism that determines the surface termination is therefore required.

Using BTO capacitors with symmetric interfaces, we investigated the FE critical thickness and experimentally confirmed a robust FE response down to $t_{BTO}$ = 3.5 u.c.. We performed PFM hysteresis loop measurements at 10×10 grid points on a capacitor with 5 × 5 µm$^2$ sizes. **Figure 3**a displays the calculated $A_{+max}/A_{-max}$ ratio of PFM loops measured at each grid point for a $t_{BTO}$ = 3.5 u.c. capacitor. The local variation of the FE response is much smaller than in the sample with mixed BTO termination. For the majority of points, the $A_{+max}/A_{-max}$ values are very close to 1.0, indicating symmetric and switchable FE polarization. The averaged PFM hysteresis loops (Figure 3b) also show symmetric butterfly-shaped amplitude-voltage curves and fully saturated phase-voltage curves that are uniform within the error bar. Template matching analysis[31] on the STEM images also shows that the Ti ions are displaced by about 0.1 Å relative to the center position of surrounding Ba ions, further supporting the polar nature of 3.5 u.c. BTO samples (Figure S7, Supporting Information). On the other hand, the FE signatures disappear for BTO capacitors with $t_{BTO}$ = 2.5 u.c. (Figure 3c), meaning the FE critical thickness in our devices is 3.5 u.c.. This value is smaller than some theoretical values[17,32] but is consistent with the DFT calculations of G. Gera *et al.*, who included charge distribution effects at the interface beyond the Thomas-Fermi capacitor.[3]

In summary, we investigated the significant effect of background oxygen pressure on SRO/BTO interface termination. We found that symmetric interfaces with uniform $TiO_2$-terminations are crucial for decreasing ferroelectric critical thickness in SRO/BTO/SRO heterostructures, and can only be obtained utilizing a lower oxygen condition (i.e. around $P_{O2}$ = 5 mTorr). As a result, we succeeded in demonstrating the theoretically predicted critical thickness of 3.5 u.c. in a real FE capacitor.[3] Our results suggest that termination control at the atomic scale will serve as a useful tool for exploring the emergent properties of oxide heterostructures and functional devices.

*Experimental*

*Sample fabrication and structural characterization*: The SRO/BTO/SRO heterostructures were grown using the PLD technique on $SrTiO_3$ (001) substrates with a miscut angle of less than 0.1°. A KrF excimer laser (248 nm, COMPex pro, Coherent) was used to ablate the SRO or BTO ceramic targets. The deposition temperature was maintained at 700 °C during the entire growth process. BTO epitaxial films were grown under $P_{O2}$ conditions of 5 mTorr and 150 mTorr. SRO top and bottom films were grown under $P_{O2}$ conditions of 100 mTorr.

The high quality of the films was confirmed using a scanning transmission electron microscope (JEM-ARM200F, JEOL). The structure of the samples was investigated by surface x-ray scattering measurements and coherent Bragg rod analysis (COBRA) performed using Huber six-circle diffractometers at Sector 12ID-D of the Advanced Photon Source (APS) and at Sector 9C of the Pohang Light Source (PLS). For the electrical measurements of the BTO layer, capacitor devices with diameters from 500 nm to 10 μm were fabricated using e-beam lithography and ion milling.

*PFM measurements*: The ferroelectric polarization switching properties were measured using an atomic force microscopy (AFM) system (Cypher Asylum) at room temperature. A commercially available Cr/Pt-coated probe tip with a spring constant of ~ 40 N/m and a resonant frequency of ~ 400 kHz (Tap300E, Budget Sensors) was used. The contact resonance frequency was 1.2 – 1.3 MHz. The high spring constant and contact resonance frequency minimized possible effects of non-piezoelectric response, such as electrostatic force.[33,34]

*Acknowledgements*
This work was supported by IBS-R009-D1 through the Research Center Program of the Institute for Basic Science in Korea. S.H.C. was supported by Basic Science Research Programs through the National Research Foundation of Korea (NRF-2015R1C1A1A01053163). Use of Advanced Photon Source was supported by the U.S. Department of Energy, Office of Science, Office of Basic Energy Sciences, under Contract No. DE-AC02-06CH11357. ((Supporting Information is available online from Wiley InterScience or from the author)).

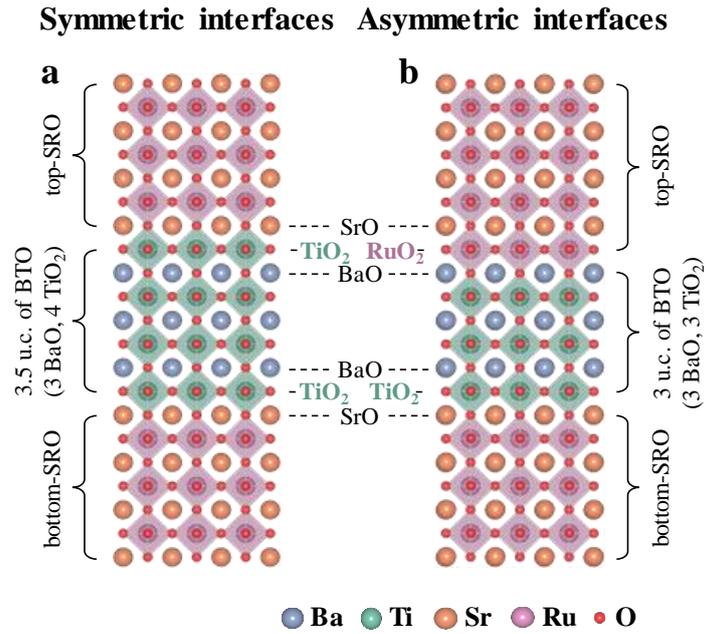

**Scheme 1.** Schematic of the possible atomic stackings of the SrRuO$_3$/BaTiO$_3$/SrRuO$_3$ (SRO/BTO/SRO) heterostructure with a) symmetric SrO-TiO$_2$ interfaces, which result in a BTO layer thickness ($t_{BTO}$) of 3.5 unit cells (u.c.), and with b) BaO-RuO$_2$ and SrO-TiO$_2$ interfaces at the top and bottom of the BTO layer, respectively, which result in $t_{BTO}$ = 3 u.c.. Note that the asymmetric case (b) occurs under the commonly assumed *unit-cell by unit-cell* growth mode.

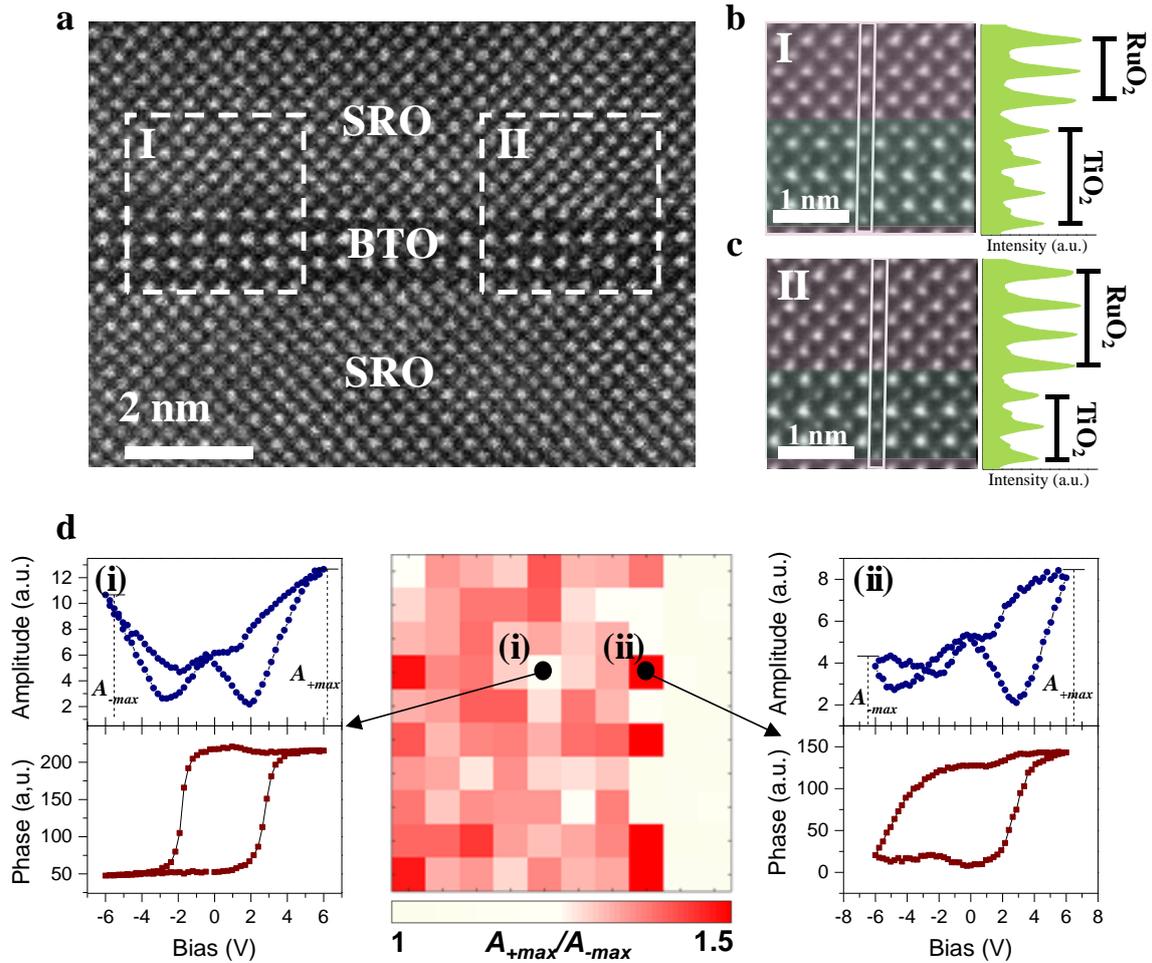

**Figure 1.** a) Scanning transmission electron microscopy (STEM) image of the SRO/BTO/SRO heterostructure with oxygen partial pressure ($P_{O2}$) during BTO growth, equal to 150 mTorr. The magnified STEM images and intensity profiles are marked by the dashed boxes in 1a and show a mixture of b) $TiO_2$ and c) BaO top terminations of the BTO layer. The BTO thicknesses are 3.5 u.c. (3 BaO and 4 $TiO_2$) and 3 u.c. (3 BaO and 3 $TiO_2$) for 1b and 1c, respectively. d) The amplitude and phase from piezoresponse force microscopy (PFM) hysteresis loops collected in different locations reveal piezoresponse in both polarization directions (left) and highly pinned piezoresponse along the downward polarization direction (right). The color map at the center of 1d displays the ratio of maximum PFM amplitude response ($A_{+max}/A_{-max}$) at positive bias ($A_{+max}$) to that at negative bias ($A_{-max}$). The black dots marked by (i) and (ii) indicate the points where the left and right plots were measured.

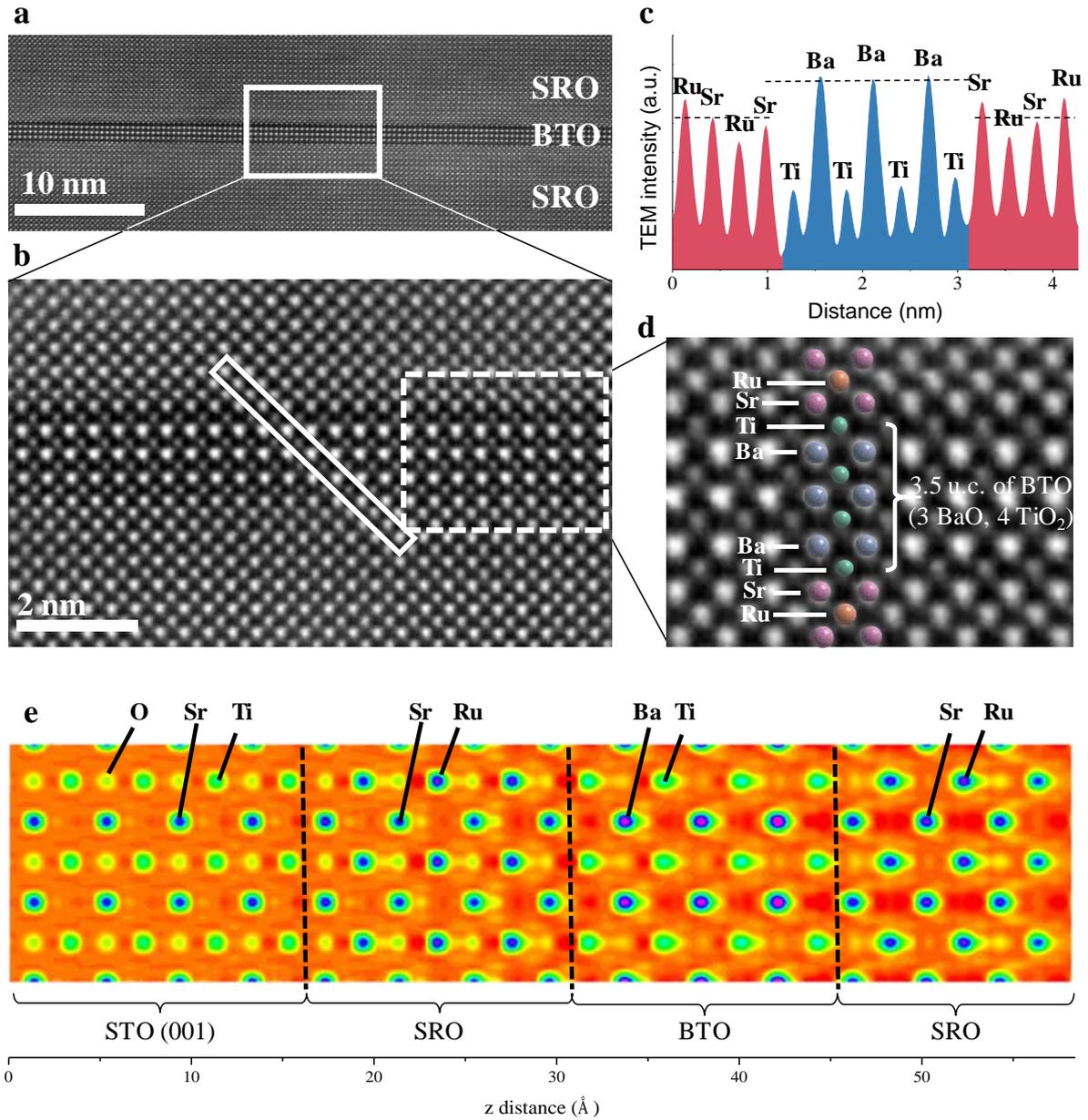

**Figure 2.** a) STEM image of the SRO/BTO/SRO heterostructure with $P_{O2}$ = 5 mTorr. b) Magnified STEM image of the region inside the white box in 2a. c) Intensity line profile corresponding to the highlighted solid box in 2b. This profile clearly shows the $TiO_2$ termination at the top and bottom of the BTO layer. d) Magnified image of the area inside the white box in 2b. The different atomic species are indicated by colored circles. In the entire measured region, the BTO layer is composed of three BaO layers and four $TiO_2$ layers, indicating $t_{BTO}$ = 3.5 u.c.. e) Coherent Bragg rod analysis (COBRA) electron density for the BTO film grown with $P_{O2}$ = 5 mTorr. The electron density map shows the (110) plane through Sr, Ba, Ti, Ru, and O. The COBRA maps show that the BTO layer has a uniform and sharp interfaces over a large length scale (i.e., at least ~ 30 μm).

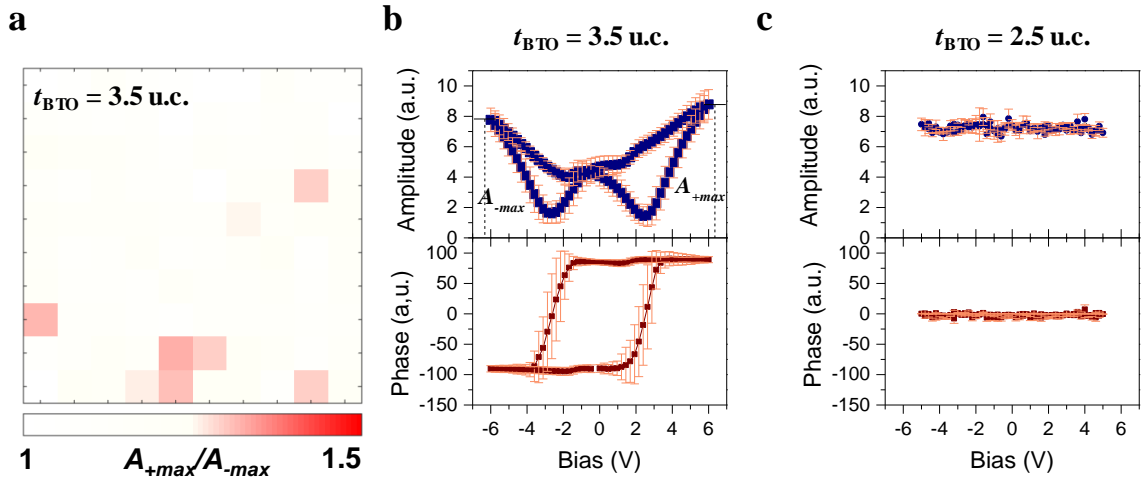

**Figure 3.** a) Color map of $A_{+max}/A_{-max}$ for the SRO/BTO/SRO capacitor with $t_{BTO}$ = 3.5 u.c. and with $P_{O2}$ = 5 mTorr. b) Averaged PFM amplitude and phase hysteresis loops measured at 10×10 different positions in a single capacitor. The error bar indicates the standard deviation of the measured loops. c) Averaged PFM amplitude and phase hysteresis loops of the SRO/BTO/SRO capacitor with $t_{BTO}$ = 2.5 u.c..

# Interface control of ferroelectricity in a SrRuO$_3$/BaTiO$_3$/SrRuO$_3$ capacitor and its critical thickness

*Yeong Jae Shin, Yoonkoo Kim, Sung-Jin Kang, Ho-Hyun Nahm, Pattukkannu Murugavel, Jeong Rae Kim, Myung Rae Cho, Lingfei Wang, Sang Mo Yang, Jong-Gul Yoon, Jin-Seok Chung, Miyoung Kim, Hua Zhou, Seo Hyoung Chang\*, Tae Won Noh\**

Y. J. Shin, Dr. S.-J. Kang, Dr. H.-H. Nahm, J. R. Kim, Dr. M. R. Cho, Dr. L. Wang, Prof. T. W. Noh
Center for Correlated Electron Systems, Institute for Basic Science (IBS), Seoul 08826, Republic of Korea
Department of Physics and Astronomy, Seoul National University, Seoul 08826, Republic of Korea
E-mail: twnoh@snu.ac.kr
Y. Kim, Prof. M. Kim
Department of Materials Science and Engineering and Research Institute of Advanced Materials, Seoul National University, Seoul 08826, Republic of Korea
Prof. P. Murugavel
Department of Physics, Indian Institute of Technology Madras, Chennai 600036, India
Prof. S. M. Yang
Department of Physics, Sookmyung Women's University, Seoul 04310, Republic of Korea
Prof. J.-G. Yoon
Department of Physics, University of Suwon, Hwaseong, Gyunggi-do 18323, Republic of Korea
Prof. J.-S. Chung
Department of Physics, Soongsil University, Seoul 06978, Republic of Korea
Dr. H. Zhou
Advanced Photon Source, Argonne National Laboratory, Lemont, Illinois 60439, United States
Prof. S. H. Chang
Department of Physics, Pukyong National University, Busan 48513, Republic of Korea
E-mail: cshyoung@pknu.ac.kr

## I. RHEED monitoring of the BTO layer growth and defining layer thickness

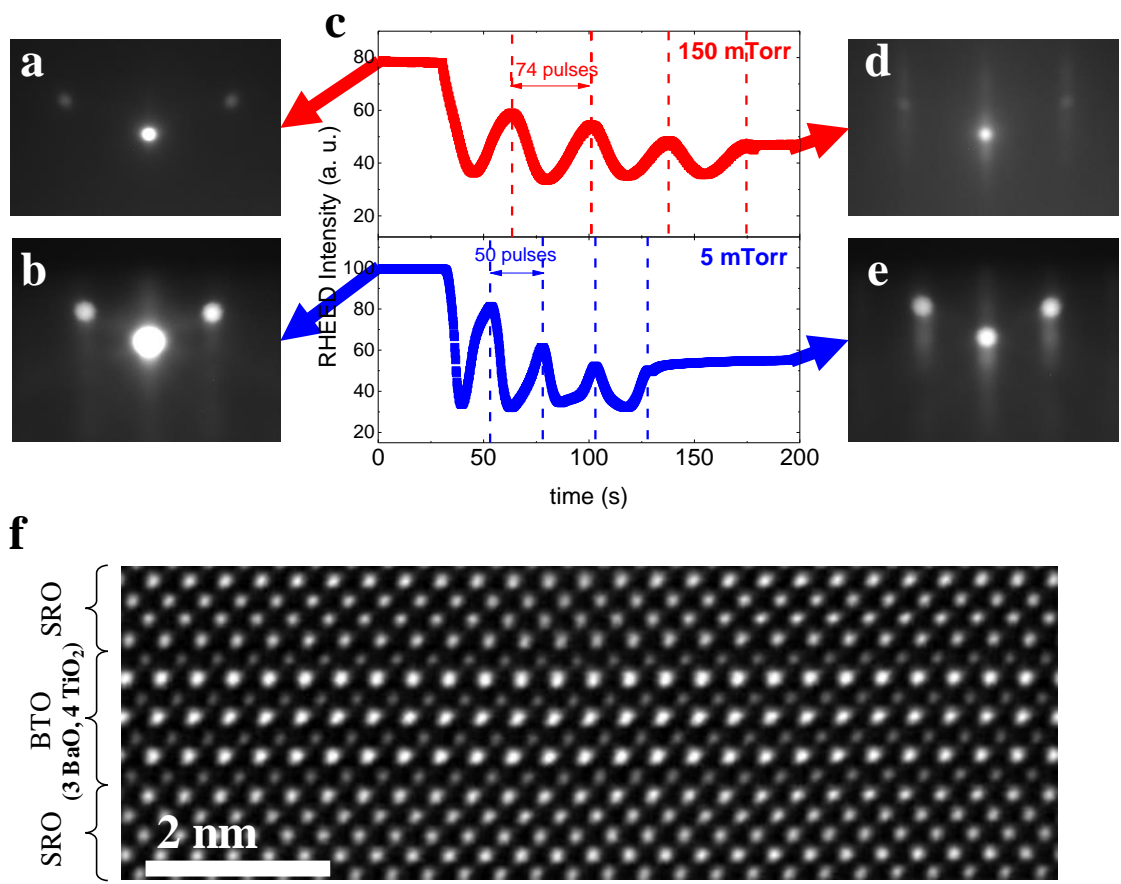

**Figure S1.** a, b) Patterns of reflective high-energy electron diffraction (RHEED) before BaTiO$_3$ (BTO) growth with oxygen partial pressure ($P_{O2}$) equal to 150 mTorr, and $P_{O2}$ = 5 mTorr, respectively. c) RHEED specular spot intensity change during BTO growth with $P_{O2}$ = 150 mTorr (red), and $P_{O2}$ = 5 mTorr (blue). d,e) RHEED patterns after BTO growth with $P_{O2}$ = 150 mTorr, and $P_{O2}$ = 5 mTorr, respectively. f) Scanning transmission electron microscopy (STEM) image of SRO/BTO/SRO heterostructure with $P_{O2}$ = 5 mTorr. Note that number of RHEED oscillations for the BTO layer is four (S1c) but the number of actual Ba layers in STEM image is only three.

The growth of BaTiO$_3$ (BTO) layers at 150 and 5 mTorr of oxygen partial pressure ($P_{O2}$) was monitored by using *in situ* reflective high energy electron diffraction (RHEED). **Figure S1**c shows the intensity oscillation of a specular spot of RHEED during the BTO layer growth at $P_{O2}$ = 150 mTorr (upper) and $P_{O2}$ = 5 mTorr (bottom). These two films were fabricated in a successive manner, with all other conditions remaining the same. Although the oxygen pressure was reduced by a factor of 30, the BTO layer grown at $P_{O2}$ = 5 mTorr still showed clear intensity oscillations, which indicates a layer-by-layer growth process. A comparison of the RHEED pattern before BTO growth (Figure S1a and S1b) and after BTO growth (Figure S1d and S1e) did not reveal surface structural changes during growth. The thickness of the BTO layer with $P_{O2}$ = 5 mTorr is measured by a scanning transmission electron microscopy (STEM) (Figure S1f). The STEM image reveals that the number ($n$) of RHEED oscillations is not consistent with the number of BTO unit cells (u.c.). Even though we intentionally make $n$ = 4 during the BTO layer growth (Figure S1c), the STEM image only exhibits three layers of Ba. In such a case, we will define the thickness of BTO layer as ($n$ - 0.5) u.c..

## II. Topography and crystal structure of the SrRuO₃/BaTiO₃/SrRuO₃ heterostructures

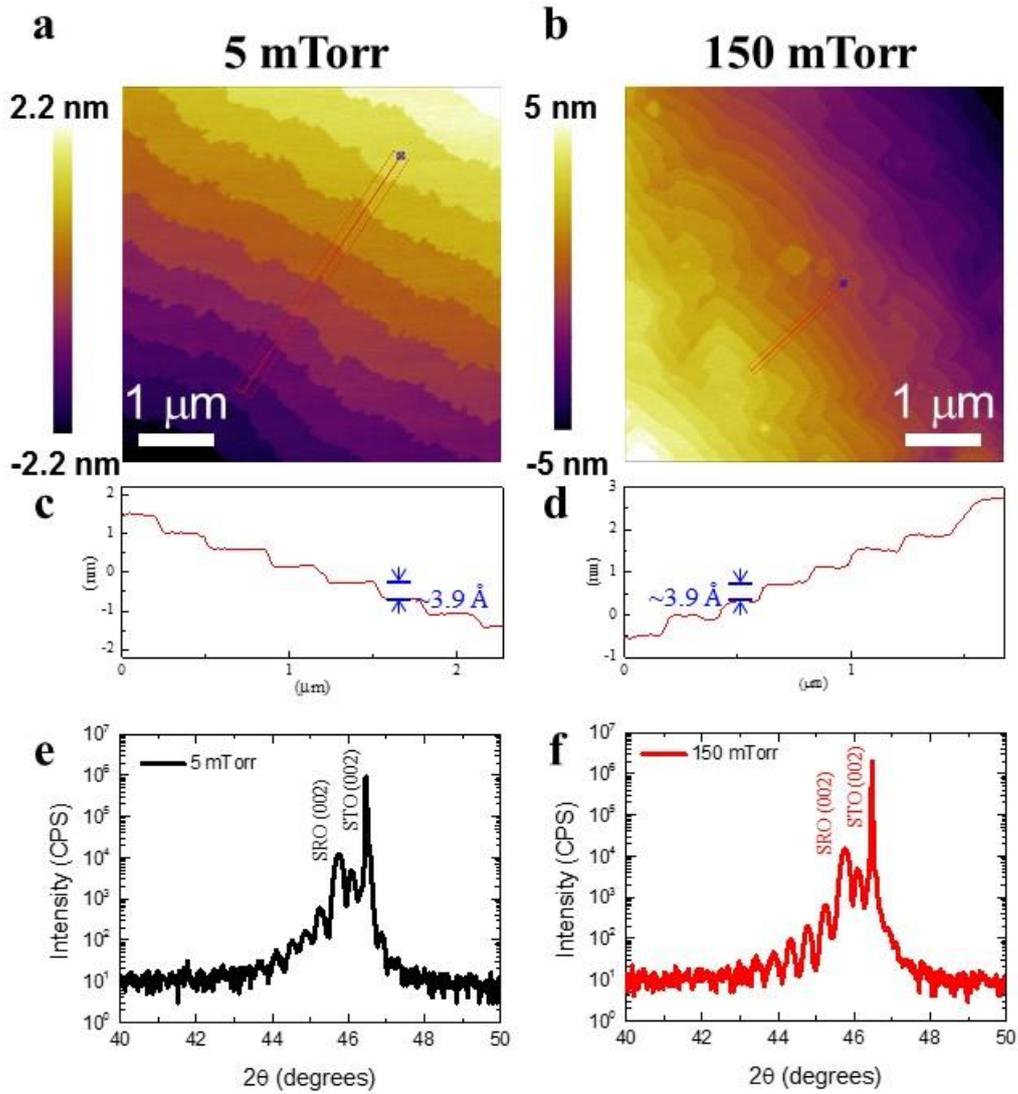

**Figure S2.** AFM topography image of the SrRuO₃/BaTiO₃/SrRuO₃ (SRO/BTO/SRO) heterostructure with BTO layer deposited at a) $P_{O2}$ = 5 mTorr and b) $P_{O2}$ = 150 mTorr. c, d) Line profiles of the highlighted regions (red line) in S2a and S2b, respectively. e,f) XRD 2$\theta$-$\theta$ scans with 20 nm top and bottom SRO electrodes and 3.5 u.c. BTO: e) $P_{O2}$ = 5 mTorr and f) $P_{O2}$ = 150 mTorr.

The surface topography of the SrRuO₃/BaTiO₃/SrRuO₃ (SRO/BTO/SRO) heterostructures was measured using atomic force microscopy (AFM). **Figure S2**a and S2b show the AFM topography image of a 3.5 u.c. BTO film with $P_{O2}$ = 5 mTorr and $P_{O2}$ = 150 mTorr, sandwiched by 20 nm SRO top and bottom electrodes. Both images show atomically flat surfaces and clear step and terrace structures. The height of the step evaluated by the line profile (Figures S2c and S2d) exactly corresponds to the height of a monolayer of perovskite oxide. Even though the BTO layer with $P_{O2}$ = 150 mTorr exhibits a half-unit-cell step height in STEM measurement, the topography image of the SRO/BTO/SRO capacitor does not show step heights of ~0.2 nm. For the case of SRO film growth, half-integer-unit-cell growth occurs at the early stage and the surface undergoes the termination layer inversion from $RuO_2$ to SrO within the growth of 1~2 monolayers.[1,2] Note that such termination layer inversion of SRO is due to the highly volatile nature of the Ru atom which prevents the formation of a $RuO_2$-terminated SRO surface. As a result, the AFM topography image of our SRO/BTO/SRO heterostructure (Figure S2b and S2d) exhibits a height difference of ~0.4 nm.

The crystal structures of the heterostructure were investigated using X-ray diffraction (XRD); the $2\theta$-$\theta$ scans of the BTO layer grown at $P_{O2}$ = 5 mTorr (Figure S2e) and $P_{O2}$ = 150 mTorr (Figure S2f) show clear Pendellosung fringes and no other phases of SRO. The peaks of the BTO layers in the different samples cannot be detected because of their thin film thicknesses (~ 1.2 nm).

## III. Detailed investigation of the mixture of interface terminations

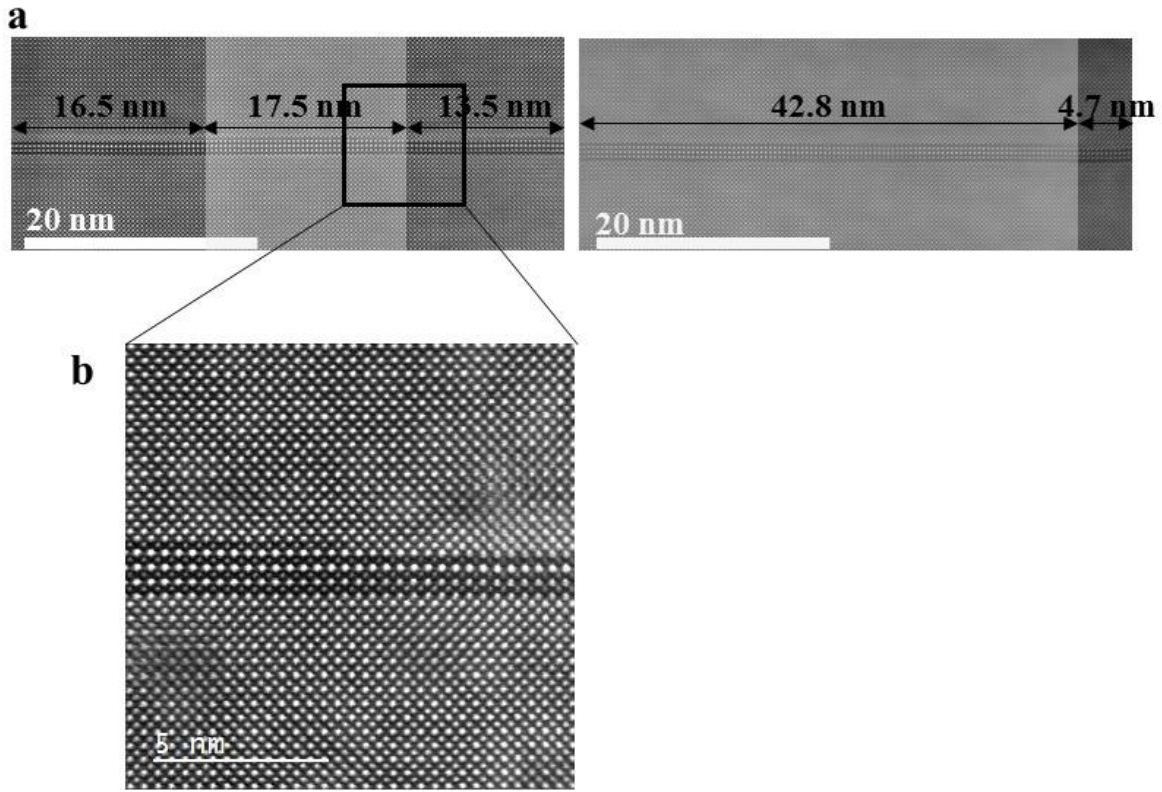

**Figure S3.** a) Average Background Subtraction-filtered (ABSF) high angle annular dark field (HAADF)-STEM images of the SRO/BTO/SRO heterostructure grown at $P_{O2}$ = 150 mTorr. The region with a TiO$_2$-terminated BTO layer is highlighted in gray. b) Magnified STEM image, which is marked by a solid box in S3a

We investigated the mixed termination layers for the BTO film grown at $P_{O2}$ = 150 mTorr using STEM. We obtained high-angle annular dark field (HAADF) images in different positions. The HAADF images taken over a scale of several μm confirmed the *half-unit-cell modulation* of the BTO layer thickness ($t_{BTO}$). A representative image is shown in **Figure S3**a. These images are taken at the boundary of the region with different terminations and clearly show the coexistence of the TiO$_2$ and BaO terminations of the BTO layer (Figure S3b). The images also exhibit a variation of the local length scale of BaO or TiO$_2$ terminated region. For example, the length of TiO$_2$-terminated BTO layer varied from ~17 nm to 40 nm or larger.

## IV. Grid-PFM measurements.

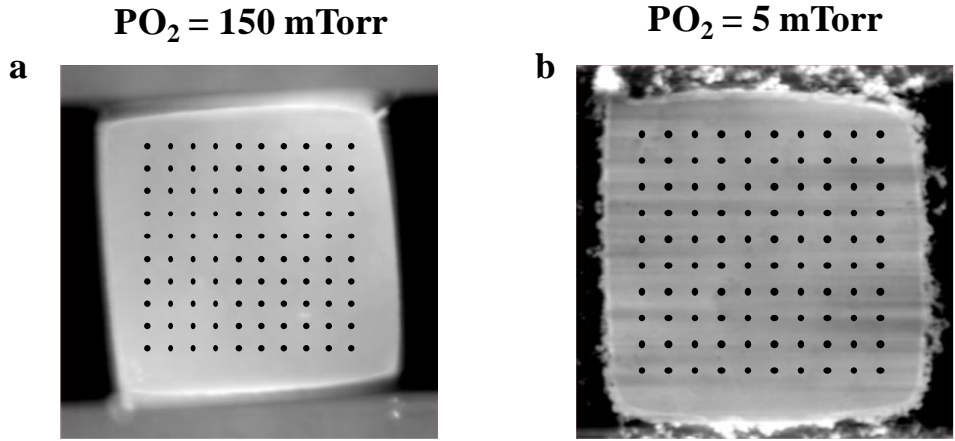

**Figure S4.** AFM topography images of 5 × 5 μm² square-shaped SRO/BTO/SRO ferroelectric capacitor with a) $P_{O2}$ = 150 mTorr and b) $P_{O2}$ = 5 mTorr. The black dots schematically show the locations at which the PFM hysteresis loops were measured.

To carefully investigate the spatial variation of the ferroelectric (FE) switching properties, we divided a single capacitor into a 10 × 10 grid and measured PFM hysteresis loops at each grid point (grid–PFM). The images of the SRO/BTO/SRO capacitor with $P_{O2}$ = 150 mTorr and $P_{O2}$ = 5 mTorr are shown in **Figure S4**a and Figure S4b, respectively. The black dots indicate the positions at which grid-PFM is performed.

## V. Cation stoichiometry of BTO films

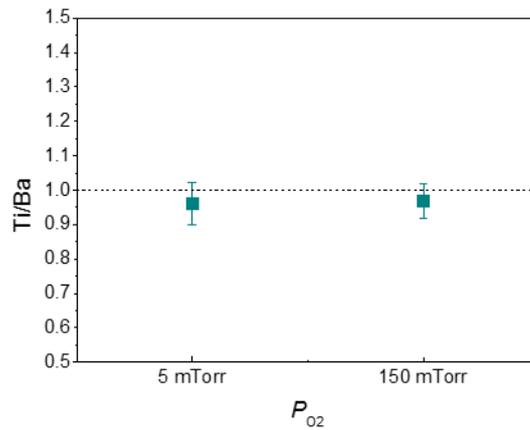

**Figure S5.** Cation Ti/Ba ratio of BTO films with different $P_{O2}$, measured by energy dispersive spectroscopy (EDS) mapping. Error bars indicate the standard deviation over different BTO film locations.

The stoichiometry of the BTO films was also investigated by energy dispersive spectroscopy (EDS). **Figure S5** shows the cation Ti/Ba ratio of the BTO films with $t_{BTO} = 20$ nm, grown at $P_{O2} = 5$ mTorr and at $P_{O2} = 150$ mTorr. In both cases, BTO is slightly Ti-poor, although the ratio is almost identical within error bars for both $P_{O2}$ conditions. These results indicate that a variation of the ablation plume stoichiometry cannot explain the interface change exhibited by our SRO/BTO/SRO heterostructures.

## VI. Theoretical calculation of the BaTiO$_3$ surface for different $P_{O_2}$ conditions.

First-principle density functional theory (DFT) calculations were performed using the Vienna Ab-initio Simulation Package (VASP).[3] Projector augmented wave (PAW) pseudopotentials[4] were used along with a plane-wave basis set with a kinetic energy cutoff of 400 eV and a 4 × 4 × 1 **k**-point mesh. The generalized gradient approximation (GGA) with the on-site $U$ (GGA+$U$) approach,[5] which is based on the Perdew-Burke-Ernzerhof (PBE)[6] exchange-correlation functional, was applied to mimic the magnetic and electronic properties of the itinerant ferromagnetic SRO metal. An effective $U$ ($U_{eff} = U$-$J$) value of 1.0 eV for the $d$ orbital was chosen, which is consistent with other relevant DFT calculations.[7]

The thermal stability of the surface termination was studied through the calculated surface Gibbs free energies of the asymmetric slabs[8,9] with 3.5 (TiO$_2$ termination) to 4 (BaO termination) BTO layers on six layers of SRO metal. All of the atomic positions (except on the fixed bottom SRO layers) were fully relaxed until the Hellmann-Feynman force on each atom was reduced to 0.01 eV/Å. Because the present thin-film growing temperature is approximately 1000 K, a paraelectric BTO phase on the fixed SRO layers was assumed. To protect the artificial interactions between the periodic slab images, the slabs were separated with a converged vacuum region of 20 Å. Here, we considered (1 × 1) bulk-like terminations to determine the different behaviors of the surface terminations according to the oxygen partial pressure.

The surface Gibbs free energy ($\Omega$) for each termination was calculated by the following equations: $\Omega(T, p) = E_{surf} - (n_{Ba}\Delta\mu_{Ba} + n_O\Delta\mu_O)$ ($E_{surf} = \Delta E_{slab} - n_{Ba}\mu_{Ba}^{bulk} + n_{Ti}\mu_{Ti}^{bulk} + n_O \mu_O^{gas}$)), where $\Delta E_{slab}$ is the relative DFT total energy with respect to the bulk BTO and the fixed base of the slab ($\Delta E_{slab} = E_{slab}^{tot} - (n_{BTO}^{bulk} \mu_{BTO}^{bulk} + E_{base})$). $n_{Ba}$, $n_{Ti}$, and $n_O$ denote the surface number of Ba, Ti, and O atoms, respectively. The quantities $\mu_{BTO}^{bulk}$, $\mu_{Ba}^{bulk}$, $\mu_{Ti}^{bulk}$, and $\mu_O^{gas}$ are the total energies of the bulk BTO, Ba metal, Ti metal, and isolated O molecule, respectively. Because the three chemical potentials for Ba, Ti, and O are bound to the bulk BTO ($\mu_{BTO}^{bulk} = (\mu_{Ti} + 2\mu_O) + (\mu_{Ba} + \mu_O)$), the values of $\Omega$ for both the TiO$_2$ and BaO terminations can be rewritten as a function of $\Delta\mu_{Ba} + \Delta\mu_O$. The temperature-dependent stability region of the BaTiO$_3$ phase was calculated by considering the formation of other phases, such as the bulk Ti/Ba metal, Ti$_2$O$_3$, TiO$_2$, BaO, and BaO$_2$, assuming that precipitation did not occur on the surfaces of the other secondary phases.

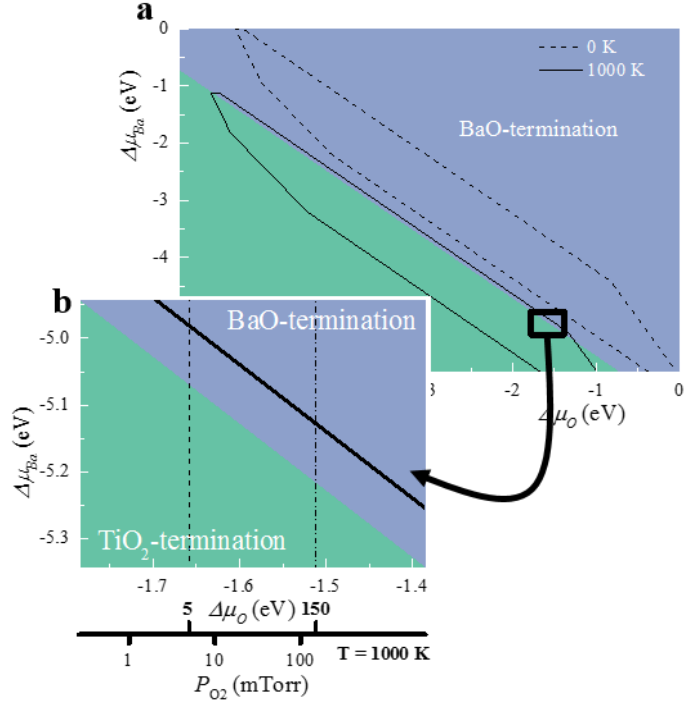

**Figure S6.** a) Calculated surface phase diagram for a paraelectric BTO surface with TiO$_2$ termination (green) and BaO termination (blue). The black dashed and solid lines show the boundaries of the stable BTO regions with respect to the secondary phases at 0 K and 1000 K, respectively. b) Enlarged phase diagram, marked by the black box in S6a. The vertical dashed and dashed-dotted lines represent the values of the oxygen chemical potentials ($\Delta\mu_O$), which correspond to the oxygen partial pressures of 5 mTorr and 150 mTorr, respectively.

The oxygen chemical potentials, $\mu_O(T, p)$, under the present growing conditions were calculated as follows: $\mu_O(T, p) = (E_{O2} + \mu_{O2}(T, p) + k_B T \ln(p/p^0))/2$, where $T$ ($T^0$) and $p$ ($p^0$) are the temperature (enthalpy reference temperature, $T^0$, = 298.15 K) and pressure (standard state pressure, $p^0$, = 1 atm), respectively. $E_{O2}$ is the calculated total energy of the O$_2$ molecule and $\mu_O(T, p)$ is the change of the oxygen chemical potential because of the temperature dependence of the enthalpy and entropy of the oxygen molecule.[10] At $T = 1000$ K, the values of $\mu_O(T, p)$ for $p = 5$ mTorr and 150 mTorr were calculated at -1.658 eV and -1.512 eV, respectively.

Based on the calculated Gibbs free energies, we determine the relative thermodynamic stability of BaO- and TiO$_2$-terminated surfaces under various environmental conditions. **Figure S6**a illustrates the phase diagram of stable termination. Here $\Delta\mu_{Ba}$ and $\Delta\mu_O$ represent the Ba- and O-chemical potentials, respectively. In the figure, the region surrounded by the dashed (solid) line indicates the stability region of a homogeneous phase at 0 K (1000 K). Outside the region, secondary phases, such as BaO$_2$ and Ti$_2$O$_3$, are formed. At 0 K, the BaO-terminated surface has a lower Gibbs free energy. As the temperature increases, the stability region moves left and enters into the TiO$_2$-terminated region. At 1000 K, which is close to our film growth temperature, both TiO$_2$ or BaO terminations are possible (Figure S6b). It is noted that decreasing $P_{O2}$ (i.e. decreasing $\Delta\mu_O$) makes TiO$_2$-terminated BTO surface become energetically more preferred at all $\Delta\mu_{Ba}$. Accordingly, when $P_{O2} = 150$ mTorr, the surface Gibbs free energies of BaO- and TiO$_2$-terminated surfaces are similar, and the entropy during growth leads to a mix-terminated surface. As we decrease $P_{O2}$ to 5 mTorr, the TiO$_2$-terminated surface is more energetically favorable, thus resulting in a uniformly terminated surface.

## VII. Evidence for ferroelectricity in SRO/BTO/SRO heterostructure with $t_{BTO} = 3.5$ u.c.

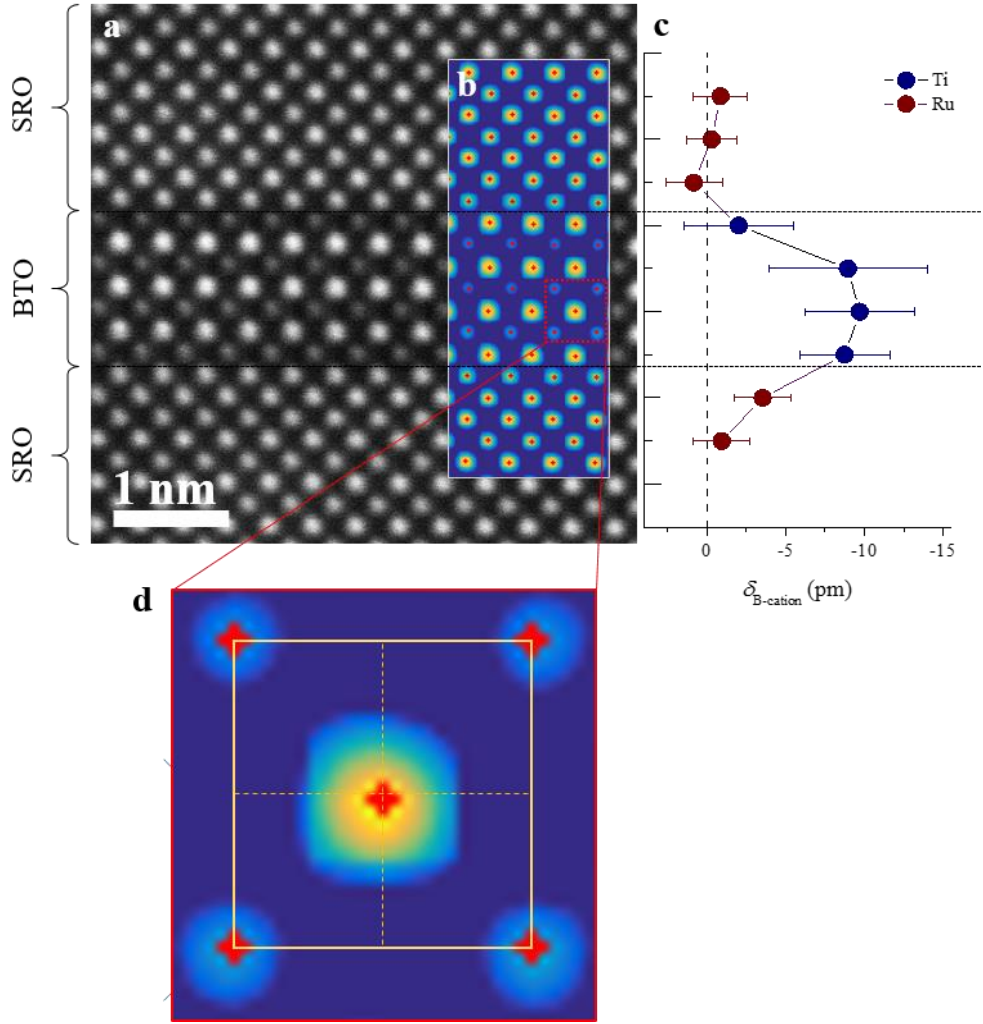

**Figure S7.** a) STEM image of a 3.5 u.c. BTO layer with $P_{O2} = 5$ mTorr, sandwiched by SRO electrodes. b) Spatially averaged section of S7a by template matching analysis (TeMA). The small red crosses indicate the peak positions of each atom. c) The closed circle indicates B-site cation displacement from the center position of the A-site cation cage ($\delta_{B\text{-cation}}$), obtained from the TeMA average peak positions of atoms for each row of S7a. Error bars indicate the standard deviation of different columns. d) Enlarged image of the highlighted red box in S7b. The yellow box indicates the Ba-cage and dashed lines indicate the cage center position. The difference between the Ti peak and the center position indicates a relative displacement of Ti in the downward direction.

The existence of FE polarization in the $t_{BTO} = 3.5$ u.c. sample was confirmed by atomic scale structural analysis. We utilized template matching analysis (TeMA) on the STEM HAADF image[11] to reduce the scan noise and scan distortion. A significant improvement of resolution is achieved when the STEM image (**Figure S7**a) is compared to TeMA image (Figure S7b).

Each atomic position can be assigned by the peak positions in the TeMA image (red crosses in Figure S7b). Figure S7c shows the evaluated B-site cation (Ti and Ru) displacements ($\delta_{\text{B-site}}$) from the center position of the A-site cation (Ba and Sr) cages. Please note that $\delta_{\text{B-site}}$ for the Ti atoms shows a maximum displacement of ~ 10 pm, which indicate BTO polarization. The Ti displacement can also be seen in the enlarged TeMA image (Figure S7d).

Supporting Information References